%% file: main.tex
\documentclass[a4paper,11pt]{article}
\usepackage{pos}
\usepackage{physics}
\usepackage{wrapfig}
\usepackage{comment}
\def\bsxi{{\boldsymbol \xi}}
\def\bsx{{\boldsymbol x}}
\def\psibar{{\bar\psi}}

\newcommand{\id}{1\!\!1}
\renewcommand{\vec}[1]{\boldsymbol{#1}}
\newcommand{\be}{\begin{equation}}
\newcommand{\ee}{\end{equation}}
\newcommand{\bea}{\begin{eqnarray}}
\newcommand{\eea}{\end{eqnarray}}

\newcommand{\MSbar}{\overline{\mbox{\rmii{MS}}}}
\newcommand{\lb}{\left\lbrace}
\newcommand{\rb}{\right\rbrace}

\newcommand{\rmi}[1]{{\mbox{\scriptsize #1}}}
\newcommand{\rmii}[1]{{\mbox{\tiny\rm{#1}}}}

\title{Non-perturbative thermal QCD at very high temperatures: computational strategy and hadronic screening masses}
\ShortTitle{Non-perturbative thermal QCD at very high temperatures}

\author*[a,b]{Leonardo Giusti}
\author*[c]{Davide Laudicina}
\author[a,b]{Matteo Bresciani}
\author[a,b]{Mattia Dalla Brida}
\author[d]{Tim Harris}
\author[b]{Michele Pepe}
\author[a,b]{Pietro Rescigno}

\affiliation[a]{University of Milano-Bicocca,
Piazza della Scienza 3, Milan, I-20126, Italy}

\affiliation[b]{INFN Milano-Bicocca,
Piazza della Scienza 3, Milan, I-20126, Italy}

\affiliation[c]{Fakultät für Physik und Astronomie, Institut für Theoretische Physik II, Ruhr-Universität Bochum,
44780 Bochum, Germany}

\affiliation[d]{Institute for Theoretical Physics, ETH Zürich,
Wolfgang-Pauli-Str. 27, 8093 Zürich, Switzerland}

\emailAdd{ leonardo.giusti@unimib.it}
\emailAdd{davide.laudicina@ruhr-uni-bochum.de}
\emailAdd{m.bresciani9@campus.unimib.it}
\emailAdd{mattia.dallabrida@unimib.it}
\emailAdd{harrist@phys.ethz.ch}
\emailAdd{michele.pepe@mib.infn.it}
\emailAdd{p.rescigno1@campus.unimib.it}

\abstract{We discuss a recently introduced strategy to study non-perturbatively thermal QCD
up to temperatures of the order of the electro-weak scale, combining step scaling
techniques and shifted boundary conditions. The former allow to renormalize the
theory for a range of scales which spans several orders of magnitude with a moderate
computational cost. Shifted boundary conditions remove the need for the zero
temperature subtraction in the Equation of State. As a consequence, the simulated
lattices do not have to accommodate two very different scales, the pion mass and
the temperature, at the very same spacing. Effective field theory arguments
guarantee that finite volume effects can be kept under control safely.
With this strategy the first computation of the hadronic screening spectrum has been carried out over more than two orders of magnitude in the temperature, from $T\sim 1$ GeV up to $\sim 160$ GeV. This study is complemented with the first quantitative computation of the baryonic screening mass at next-to-leading order in the three-dimensional effective theory describing QCD at high temperatures. Both for the mesonic and the baryonic screening masses, the known leading behaviour in the coupling constant is found to be not sufficient to explain the non-perturbative data over the entire range of temperatures. These findings shed further light on the limited applicability of the perturbative approach at finite temperature, even at the electro-weak scale.}

\FullConference{The 41st International Symposium on Lattice Field Theory (LATTICE2024)\\
 28 July - 3 August 2024\\
Liverpool, UK\\}

\begin{document}
\maketitle

\section{Introduction}
\input{introduction.tex}

\section{Preliminaries on the effective theory}
\input{preliminaries.tex}

\section{Non-perturbative thermal QCD at very high temperatures}
\input{NPT_QCD.tex}

\section{Correlation functions and screening masses}
\input{Corr.tex}
\section{Numerical results}
\input{Num.tex}
\section{Conclusions}
\input{conclusion.tex}
\input{ack}
\appendix
\section{Baryonic thermal screening mass at NLO}
\input{appA.tex}
\bibliographystyle{JHEP}
\bibliography{bibfile}

\end{document}

%% file: introduction.tex
Quantum Chromodynamics (QCD) at very high temperatures plays a pivotal r\^ole in particle and nuclear physics as well as in cosmology. At asymptotically high temperatures, thermal QCD is described by a three-dimensional effective gauge theory, whose dynamics is non-perturbative ~\cite{Ginsparg:1980ef,Appelquist:1981vg}. The study of such an effective theory from a perturbative point of view is then limited by the so-called infrared problem, which manifests itself when ultrasoft gluons enter into loops ~\cite{Linde:1980ts}. This implies that perturbation theory can predict the coefficients of the expansion in the strong coupling constant $g$ only up to a finite order and, as a consequence, that the theory has to be solved non-perturbatively even at very high temperatures. An important example is provided by the Equation of State, where non-perturbative contributions start at $O(g^6)$ and are still relevant at very high temperatures\cite{Braaten:1995jr,PhysRevD.67.105008}. These findings clearly point out that, in order to have a reliable and satisfactory understanding of the dynamics of the high temperature regime of QCD, a fully non-perturbative approach is essential up to temperatures which are, at least, of the order of magnitude of electro-weak scale. The strategy that we outline here provides a solid framework to achieve such a non-perturbative description. First introduced for the Yang-Mills theory in Ref. \cite{Giusti:2016iqr}, and then generalized to QCD in Ref. \cite{DallaBrida:2021ddx}, this strategy allows to simulate the theory of strong interactions up to very high temperatures from first principles with a moderate computational effort.

As a concrete application we report the results that we obtained in the calculation of the hadronic screening masses, which were carried out in Refs. \cite{DallaBrida:2021ddx, Giusti:2024ohu}, in presence of $N_f=3$ massless quarks in a temperature interval ranging from $T=1$ GeV up to $\sim 160$ GeV. Those observables probe the exponential
fall-off of two-point correlation functions of hadronic interpolating operators in the spatial directions and are the inverses of spatial correlation lengths, which characterize the response of the plasma when hadrons are injected into it. Their $O(g^2)$ contribution is known since a long time for the mesonic sector \cite{Laine:2003bd}, and has been recently calculated for the baryonic one \cite{Giusti:2024mwq}. For this reason, beyond their intrinsic theoretical interest, they also provide a further check of the reliability of perturbation theory up to very high temperatures.

%% file: preliminaries.tex
\label{sec:eft}
In QCD at very high temperatures field fluctuations at energies which are much larger than the temperature decouple and the resulting theory is effectively three-dimensional with field content given by zero Matsubara gauge modes only.

\subsection{Gluonic sector}
The gluonic sector of the effective theory contains
gauge fields $A_k^{ }$, with $k=1,2,3$, living in three spatial
dimensions, whose dynamics
is non-perturbative \cite{Linde:1980ts}. They are coupled to a
massive scalar field $A_0^{ }$, which transforms under the adjoint
representation of the gauge group. 
By taking into account these degrees of freedom,
the corresponding effective action, which is usually called
Electrostatic QCD (EQCD), reads
\cite{Ginsparg:1980ef,Appelquist:1981vg}
\begin{align}
\label{eq:EQCD}
    S_\rmii{EQCD}^{ } \, = \, \int\! {\rm d}^3_{ }x\,
   \lb \frac{1}{2}
   \Tr \left[ F_{ij}^{ }F_{ij}^{ } \right] 
  +\Tr\left[\left(D_j^{ } A_0^{ } \right)\left(D_j^{ } A_0^{ } \right) \right]
  + m_\rmii{E}^2 \Tr \left[ A_0^2 \right]\rb \, + \, \dots \, ,
\end{align}
where the dots stand for higher-dimensional operators
\cite{Laine:2016hma}. Here $i,j=1,2,3$ and 
$
 [ D_i^{ },D_j^{ } ]=-i g_\rmii{E}^{ } F_{ij}^{ }
$ with the covariant derivative defined as
$
 D_i^{ }=\partial_i^{ }-ig_\rmii{E}^{ }A_i^{ }
$.  
The matching coefficients 
$m^2_\rmii{E}$ and $g^2_\rmii{E}$ parametrize the Lagrangian
mass squared of the scalar field $A_0^{ }$ and
the three-dimensional coupling constant respectively. For $N_f=3$, at leading order in perturbation theory, in terms of the QCD coupling $g$, they read $m^2_\rmii{E}=\frac{3}{2}g^2 T^2$ and $g^2_\rmii{E}=g^2 T$ respectively \cite{Kapusta:1979fh,Laine:2005ai,Ghisoiu:2015uza}. At asymptotically high $T$, the coupling $g$ is small and three different energy
scales develop so that 
\be
\frac{g^2_{_{\rmii E}}}{\pi} \ll m_{_{\rmii E}} \ll \pi T\; .  
\ee
If one is interested in processes at scales of $O(g^2_{_{\rm E}})$, the scalar field $A_0$ can be integrated out.
The action of the remaining effective theory, dubbed Magnetostatic QCD (MQCD), is given by
\be
S_{\rmii{MQCD}} = \frac{1}{g^2_{_{\rmii E}}} \int d^3 x \Big\{\frac{1}{2} \Tr\left[F_{ij} F_{ij}\right] \Big\} +\dots
\ee
This is a three-dimensional Yang--Mills theory, it has a non-perturbative dynamics
and therefore it needs to be solved non-perturbatively~\cite{Linde:1980ts}. Being $g_{\rmii E}^2$ the only dimensionful scale of the theory, one would expect all
dimensionful quantities to be proportional to the appropriate power of $g^2_{_{\rm E}}$ times a non-perturbative coefficient.

\subsection{Fermionic sector}
At variance of the gluonic fields, at high temperatures quarks are always
heavy fields with masses $\sim \pi T$, due to fermionic Matsubara
frequencies. Therefore, the dynamics of such fields is described by
a three-dimensional non-relativistic QCD 
(NRQCD) action, ~\cite{Huang:1995tz,Caswell:1985ui,Brambilla:1999xf}, which reads at $O(g_{\rm E}^2/(\pi T))$ for $N_\rmi{\sl f}^{ }=3$ massless
fermions in the lowest Matsubara sectors
\begin{align}
\label{eq:NRQCD}
    \begin{aligned}
        S_\rmii{NRQCD}  = i \sum_\rmi{\sl f\,=\it u,d,s}
 \int\! {\rm d}^3_{ } x \, \biggl\{
 & \bar{\chi}_\rmi{\sl f\,}^{ }(x) 
 \left[ M-g_\rmii{E}^{ } A_0^{ } +D_3^{ } 
 -\frac{\nabla_\perp^2}{2 \pi T}\right]\chi_\rmi{\sl f\,}^{ }(x) \\
 -& \bar{\phi}_\rmi{\sl f\,}^{ }(x) \left[ M+g_\rmii{E}^{ } A_0^{ } +D_3^{ }
 -\frac{\nabla_\perp^2}{2 \pi T} \right]\phi_\rmi{\sl f\,}^{ }(x) \biggr\}
 \, + O\left(\frac{g_{\rmii E}^2}{\pi T}\right) .
    \end{aligned}
\end{align}
Here the low energy constant is $M \, = \, \pi T \left[1+g^2/(6\pi^2)\right]$ ~\cite{Laine:2003bd} and $\chi$ and $\phi$ are two-component Weyl spinors, whose expression in the lowest Matsubara sector reads
\begin{align}
\label{eq:chiphi}
    \psi_{f} (x_0,x) = \sqrt{T} e^{i\pi T x_0}
    \begin{pmatrix}
        \chi_f(x)\\
        \phi_f(x)
    \end{pmatrix} \,,
\end{align}
where $\psi$ is the usual four dimensional fermion field and $f$ is a flavour index, see appendix A of Ref. \cite{Giusti:2024mwq} for a further discussion.
Therefore the QCD dynamics at high temperature whose field content is given by fermion fields in the lowest fermionic Matsubara sector, which interact with soft ($A_0$) and ultrasoft ($A_k$) gauge modes only, is described by $S_{\rmii{QCD}_3} = S_\rmii{EQCD} + S_\rmii{NRQCD}$.

%% file: NPT_QCD.tex
\label{sec:strategy}
\subsection{Renormalization and lines of constant physics}
A hadronic scheme is not a convenient choice to renormalize QCD non-perturbatively when
considering a broad range of temperatures spanning several orders of magnitude. This
would require to accommodate on a single lattice both the temperature and
the hadronic scale which may differ by orders of magnitude, making the numerical computation extremely challenging. A similar problem is encountered when renormalizing QCD
non-perturbatively, and it was solved many years ago by introducing a
step scaling technique~\cite{Luscher:1991wu,Jansen:1995ck}. 

In order to solve this problem, we build on that knowledge by considering a non-perturbative definition
of the coupling constant in a finite volume, $\bar g^2_{\rm SF}(\mu)$, which can be computed precisely on the lattice for values
of the renormalization scale $\mu$ which span several orders of magnitude. Making a definite
choice, in this section we use the definition based on the Schr\"odinger functional (SF)~\cite{Luscher:1993gh}, however, notice that, other possible choices are available. In particular, in our lattice setup we also made use of the gradient flow (GF) definition of the running coupling ~\cite{Fritzsch:2013hda,Brida:2016flw,DallaBrida:2016kgh}, see appendix B of Ref. \cite{DallaBrida:2021ddx}. Once $\bar g^2_{\rm SF}(\mu)$ is known in the
continuum limit for $\mu \sim T$ \cite{Brida:2016flw,DallaBrida:2018rfy}, thermal QCD can be renormalized by fixing the value of the running coupling constant at fixed lattice spacing $a$ to be
\begin{align}
    \bar g^2_{\rm SF}(g_0^2, a\mu) = \bar g^2_{\rm SF}(\mu)\; ,\qquad a\mu\ll 1\;.
\end{align}
This is the condition, together with the definition of the critical mass, see App. B of Ref. \cite{DallaBrida:2021ddx}, that fixes the so-called lines of constant physics, i.e. the dependence of
the bare coupling constant $g_0^2$ and of the quark mass on the lattice spacing, for values of $a$ at which the scale $\mu$ and
therefore the temperature $T$ can be easily accommodated. For a more complete discussion on how this technique is implemented in practical lattice simulations we refer to appendix B of Ref. \cite{DallaBrida:2021ddx}.
\subsection{Shifted boundary conditions}
\label{sec:sbc}
The thermal theory is defined by requiring that the fields satisfy shifted boundary
conditions in the compact direction~\cite{Giusti:2011kt,Giusti:2010bb,Giusti:2012yj},
while we set periodic boundary conditions in the spatial directions. The former consist
in shifting the fields by the spatial vector $L_0\, \bsxi$ when crossing the boundary of the
compact direction, with the fermions having in addition the usual sign flip. For the gauge
fields they read
\begin{equation} \label{eq:shift_gluons}
  U_\mu(x_0+L_0,\bsx)= U_\mu(x_0,\bsx-L_0\bsxi)\; ,
  \quad
  U_\mu(x_0,\bsx+\hat{k}L_k)= U_\mu(x_0,\bsx)\; , 
\end{equation}
while those for the quark and the anti-quark fields are given by
\begin{align}
&\psi(x_0+L_0,\bsx)  =  -\psi(x_0,\bsx - L_0\bsxi)\; ,
\quad
& \psi(x_0,\bsx+\hat{k} L_k)  = \psi(x_0,\bsx)\; ,
\nonumber\\
&\psibar(x_0+L_0,\bsx)  =  -\psibar(x_0,\bsx - L_0\bsxi)\;,
\quad
&\psibar(x_0,\bsx+\hat{k} L_k)  = \psibar(x_0,\bsx)\; ,\label{eq:shift_quark}
\end{align}
where $L_0$ and $L_k$ are the lattice extent in the 0 and $k$-directions respectively. In the thermodynamic limit, a relativistic thermal field theory in the presence of shift $\bsxi$ is equivalent to the 
very same theory with usual periodic (anti-periodic for fermions) boundary conditions but with a longer extension of the
compact direction by a factor $\sqrt{1+\vec \xi^2}$~\cite{Giusti:2012yj}, i.e. the standard relation between
the length and the temperature is modified as $T=1/(L_0 \sqrt{1+\vec \xi^2})$. Shifted boundary conditions
represent a very efficient setup to tackle several problems that are otherwise very challenging both from the
theoretical and the numerical viewpoint. Some recent examples are provided by the SU(3) Yang-Mills theory Equation of State which was obtained at
the permille level up to very high temperatures~\cite{Giusti:2014ila,Giusti:2016iqr} and more recently in $N_f=3$ QCD with a novel computation of the renormalization constant of the flavour-singlet local vector current \cite{Bresciani:2022lqc}. The same setup is currently in use to carry out the first non-perturbative computation of the Equation of State at large temperatures in thermal QCD \cite{DallaBrida:2020gux,Bresciani2024}. As a further remark, notice that the use of shifted boundary conditions is not
crucial for the numerical calculation presented in section \ref{sec:num}, however we have chosen to use them with $\vec{\xi}=(1,0,0)$ so as to share the cost of generating the gauge configurations with that project. Moreover, the free case computation of both the mesonic and the baryonic screening masses, reported in appendices F of Ref. \cite{DallaBrida:2021ddx} and D of Ref. \cite{Giusti:2024ohu} respectively, indicates that the use of shifted boundary conditions with $\vec{\xi}=(1,0,0)$ guarantees the lattice measurements to be affected by milder discretization errors.
\subsection{Finite-volume effects}
\label{sec:FV}
As we have seen in section~\ref{sec:eft}, at asymptotically high temperatures, the only dimensionful quantity is the three dimensional gauge coupling $g_{\rmii E}^2$. This implies that the mass gap developed by
thermal QCD must be proportional to $g^2_{_{\rm E}}$~\cite{Arnold:1995bh}. On the other hand, at intermediate temperatures, provided that the temperature is sufficiently large with respect to $\Lambda_{\rm QCD}$, the mass gap of the theory is
always expected to be proportional to the temperature times an appropriate power of the coupling
constant~\cite{Laine:2009dh}.
As a consequence, finite-size effects are exponentially suppressed with $LT$ times a coefficient that tends to decrease logarithmically
with the temperature, see Refs.~\cite{Meyer:2009kn,Giusti:2012yj}. For instance, by taking into account a generic spatial correlation function $C_{\cal O}(x_3)$, see section \ref{sec:2pt}, and by defining the finite volume residue due to the compactification in the 1-direction as
\begin{align}
    {\cal I}_1(x_3,L)\equiv \left[ 1-\lim_{x_3\to \infty} \right] C_{\cal O}(x_3)
\end{align}
by employing the transfer-matrix representation of such a correlation function, it is easy to see that, by neglecting exponentially suppressed terms, the residue can be written, in the large $x_3$ limit, as
\begin{align}
\label{eq:I1}
    {\cal I}_1(x_3,L) \overset{x_3\to\infty}{\propto} e^{-L\gamma_1(E_n+i\xi_1\omega_{n})}
\end{align}
where $\gamma_1$ and $\xi_1$ are the Lorentz factor and the shift parameter due to shifted boundary conditions respectively, see section \ref{sec:sbc}, $\omega_n$ are fermionic or bosonic Matsubara frequencies, depending on the spin quantum number of the interpolating operator $\cal O$ and $E_n$ is the energy of the lightest 1-particle state with the correct quantum numbers. It is now crucial to notice that the 1-particle energies are confined to the range $M_{\rmii{gap}} \leq E_n \leq \pi T $ and the residue ${\cal I}_1$ is then exponentially suppressed as $M_{\rmii{gap}}L\to \infty$ or equivalently for $LT\to \infty$. 
For this reason, in our lattice setup we always employ large spatial extents, i.e. $L/a=288$, so that $LT$ ranges always from $20$ to $50$.

\subsection{Restricting to the zero-topological sector}
At high temperature, the topological charge distribution is expected to be highly peaked at zero. For QCD with three light degenerate flavours of mass $m$, the dilute instanton gas approximation predicts for the the topological susceptibility  $\chi\propto m^3T^{-b}$ with $b\sim 8$.
The analogous prediction for the Yang--Mills theory has been verified explicitly on the
lattice~\cite{Giusti:2018cmp}. Similarly, computations performed in QCD seem to confirm the $T$-dependence predicted by the
semi-classical analysis even though the systematics due to the introduction of dynamical fermions is still difficult to control ~\cite{Bonati:2015vqz,Borsanyi:2016ksw}. As a result, already at low temperatures, namely at $T\sim1$ GeV, the probability to
encounter a configuration with non-zero topology in volumes large enough to keep finite volume effects under control is expected to be several orders of magnitude smaller than the permille or so. For these reasons, we can safely restrict our calculations to the sector with zero topology.

%% file: Corr.tex
\label{sec:2pt}
\subsection{Mesonic interpolating operators}
We are interested in flavour non-singlet fermionic bilinear operators of the form
\begin{equation}
\label{eq:bils}
{\cal O}^a(x) = \psibar(x) \Gamma_{{\cal O}} \, T^a \,\psi(x)\;,\\
\end{equation}
where $\Gamma_{{\cal O}}=\left\{\id,\gamma_5,\gamma_\mu,\gamma_\mu\gamma_5\right\}$ characterizes the structure of the operators in the Dirac space, with the latter named as usual as ${\cal O}=\left\{S,P,V_\mu,A_\mu\right\}$,
and we restrict ourselves to $\mu=2$. Here $a$ is a flavour index which dictates the flavour structure of the corresponding states. The associated two-point correlation function is
\begin{align}
    C_{\cal O} (x_3) =  \int dx_0 dx_1 dx_2 \expval{{\cal O}^a(x) {\cal O}^a (0)} \overset{x_3\to \infty}{\propto} e^{-m_{\cal O} x_3}
\end{align}
where $m_{\cal O}$ is the corresponding screening mass and no summation over $a$ is understood.

At zero temperature, due to the breaking of chiral symmetry, these masses are all different. However, when the temperature is large enough, in the chiral limit the vector and axial vector screening masses are
expected to become degenerate thanks to the restoration of the non-singlet chiral symmetry. Moreover, as previously mentioned, at high temperature only the sector with zero topology
contributes de facto to the
functional integral~\cite{Gross:1980br}. This leads to the degeneracy of the non-singlet scalar
and pseudoscalar screening masses as well. For a detailed investigation of such a degeneracy pattern in the same range of temperatures, see Ref. \cite{Laudicina:2022ghk}.

\subsubsection{Leading interacting contribution in the effective theory}
The $O(g^2)$ contribution to the non-singlet mesonic screening masses has been computed in the
effective theory~\cite{Laine:2003bd}. For three massless quarks, the expression reads
\begin{align}
\label{eq:meson_pt}
m^{\rm PT}_{{\cal O}} = 2 \pi T\, \left( 1+0.032739961\cdot g^2\right)\; , 
\end{align}
where the first term comes from the free field theory, while the second one is generated by the interactions. The next-to-leading order contribution in eq.~(\ref{eq:meson_pt}) is independent of the specific mesonic operator ${\cal O}$ and spin-dependent effects are expected to appear at $O(g^4)$ in the strong coupling constant \cite{Koch:1992nx,Hansson:1991kb}.

\subsection{Baryonic interpolating operators}
The simplest fermionic operator with positive, or negative parity, which carries the nucleon quantum numbers is\footnote{In order to avoid clutter we omit the free Dirac index on the rightmost $d$ quark.}
\begin{align}
\label{eq:nucleon_op}
    N(x) =  \epsilon^{abc}\left( u^{aT}(x) C\gamma_5 d^b (x) \right) d^c (x) \,,
\end{align}
where the transposition acts on the spinor indices, latin letters refer to colour indices and $C=i\gamma_0 \gamma_2$ is the charge-conjugation matrix. The contraction with the totally anti-symmetric tensor $\epsilon^{abc}$ guarantees the operator to be a colour singlet and gauge invariant.

Notice that, at variance of the mesonic case, the corresponding two-point correlation function cannot be projected to zero Matsubara frequency, since fermionic Matsubara frequencies are always non-vanishing. Therefore, the screening correlator for a nucleon interpolating operator in the lowest fermionic Matsubara sector reads
\begin{align}
\label{eq:2pt-baryon}
    C_{N^\pm} (x_3) = \int dx_0 dx_1 dx_2 e^{-i\frac{x_0+\xi x_1}{L_0}\gamma^2\pi} \expval{\Tr \left[ P_\pm N(x) \overline{N}(0) \right]} \overset{x_3\to \infty}{\propto} e^{-m_{N^\pm} x_3}\,.
\end{align}
Here $P^\pm = (1\pm \gamma_3)/2$ are projectors on positive ($N^+$) and negative ($N^-$) $x_3$-parity states respectively. For completeness, notice that the two-point function in eq. \eqref{eq:2pt-baryon} is written in presence of shifted periodic boundary conditions with $\bsxi=(\xi,0,0)$ and the usual correlator with standard periodic boundary conditions is easily recovered by setting $\xi=0$.

Similarly to the mesonic case, since at high temperatures chiral symmetry is not spontaneously broken, the positive and the negative parity two-point correlation functions are equal up to a sign in the chiral limit and, as a consequence, $m_{N^+} = m_{N^-}$. This is at variance of the zero temperature case, where, due to the spontaneous breaking of chiral symmetry, the nucleon and the $N(1535)$ masses differ by several hundreds of MeV. 
\subsubsection{Leading interacting contribution in the effective theory}
At variance of the mesonic case, so far in the literature, the only next-to-leading calculation on the baryonic screening masses was just qualitative \cite{Hansson:1994nb} and only very recently the first quantitative computation of such a perturbative correction has been carried out \cite{Giusti:2024ohu}, see appendix \ref{app:A} for the calculation. At $O(g^2)$ in the coupling constant, the baryonic screening masses, for three massless quarks, read
\begin{align}
\label{eq:baryon_pt}
    m_{N^{\pm}}^{\rm PT} = 3\pi T \left( 1+ 0.046 \cdot g^2 \right) \,,
\end{align}
where the free theory value is $3\pi T$, while the $O(g^2)$ correction is due to interactions. Notice that, at this order in perturbation theory, the positive and the negative parity partners are degenerate.

%% file: Num.tex
\label{sec:num}
As a concrete application of the strategy outlined in section \ref{sec:strategy}, in Refs \cite{DallaBrida:2021ddx,Giusti:2024ohu} we have performed numerical simulations at 12 values of the temperature, $T_0, \ldots$, $T_{11}$ covering the range from approximately $1$ GeV up to about $160$~GeV. For the 9
highest ones, $T_0, \ldots$, $T_8$, gluons are regularized with the Wilson plaquette action, while for the 3 lowest temperatures, $T_9$, $T_{10}$ and
$T_{11}$, we adopt the tree-level improved L\"uscher-Weisz gauge action. The three massless
flavours are always discretized by the $O(a)$-improved Wilson--Dirac operator.
In order to extrapolate the results to the continuum limit, several lattice spacings are simulated at each temperature with the extension of the compact dimension being $L_0/a=4,6,8$ or $10$.

\subsection{Mesonic screening masses}
\begin{figure}
    \centering
    \includegraphics[width=0.4\textwidth]{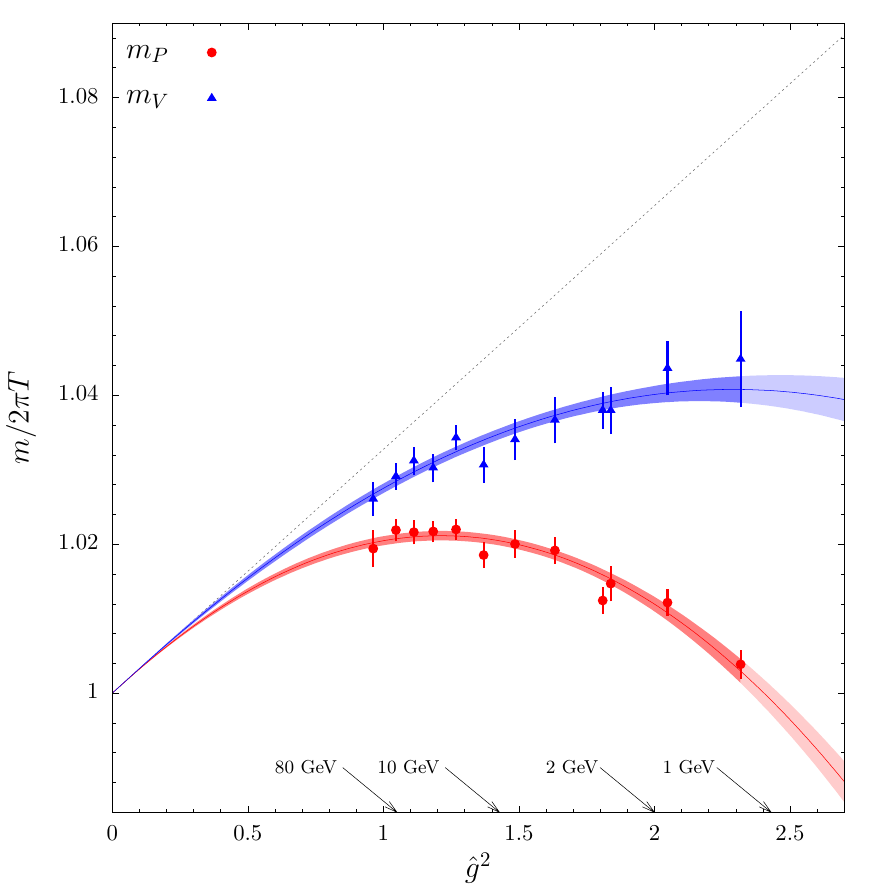} 
    \includegraphics[width=0.4\textwidth]{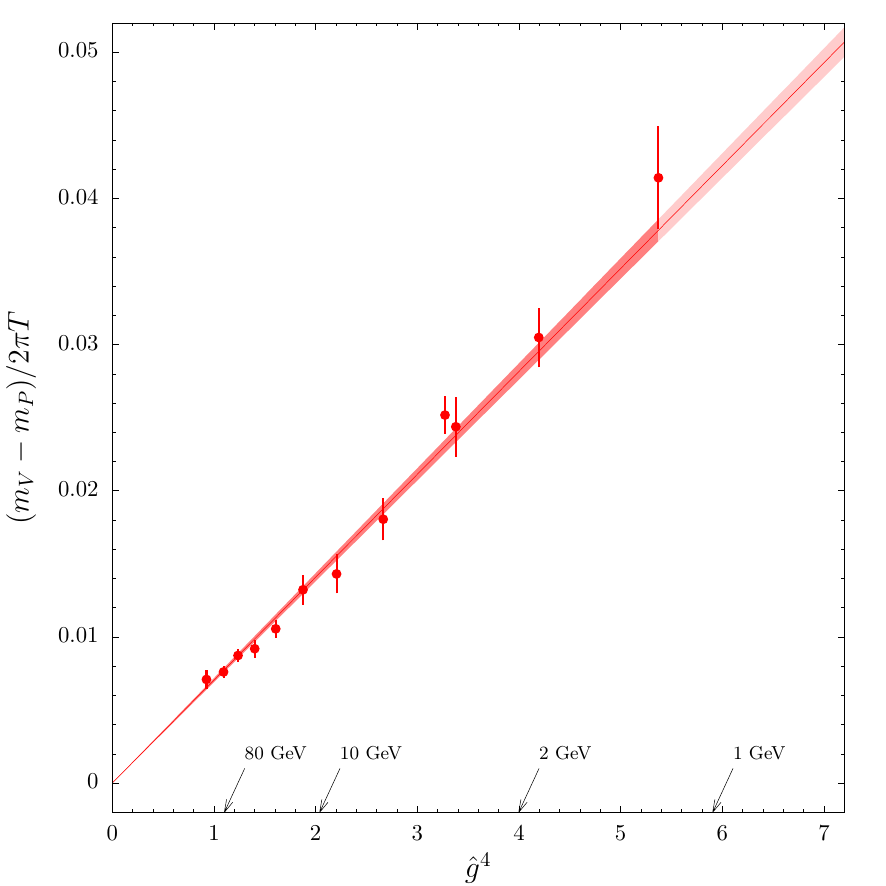} 
    \caption{Left: pseudoscalar (red) and vector (blue) screening masses versus $\hat g^2$. The bands represent the best fits in eqs.~(\ref{eq:quartic}) and (\ref{eq:mvfit}), while the dashed line is the analytically known contribution.
    Right: the vector-pseudoscalar mass difference, normalized to $2\pi T$, versus $\hat g^4$. Red bands
    represent the best fits of the data as explained in the text.\label{fig:massCL}}
\end{figure}
\label{sec:meson}
The mesonic screening masses have been computed with a few permille accuracy in the continuum limit. Within our statistical precision, the screening masses associated to the pseudoscalar and the scalar density are found to be degenerate and a similar discussion holds for the screening masses of the vector and the axial current, as expected from chiral symmetry restoration. Given the high accuracy of our non-perturbative data it has been possible to parameterize the temperature dependence of the masses. In order to do that, we introduce the function $\hat g^2 (T)$ defined as 
\begin{equation}\label{eq:gmu}
  \frac{1}{\hat g^2(T)} \equiv \frac{9}{8\pi^2} \ln  \frac{2\pi T}{\Lambda_{\MSbar}} 
  + \frac{4}{9 \pi^2} \ln \left( 2 \ln  \frac{2 \pi T}{\Lambda_{\MSbar}}  \right)\; , 
\end{equation}
where $\Lambda_{\MSbar} = 341$~MeV is taken from Ref.~\cite{Bruno:2017gxd}. It corresponds
to the 2-loop definition of the strong coupling constant in the
$\overline{\rm{MS}}$ scheme at the renormalization scale $\mu=2\pi T$. For our purposes, however, this is just a function
of the temperature $T$, dictated by the effective theory analysis, that we use to analyze our results and which makes it easier to compare with the known perturbative results.
\subsubsection{Pseudoscalar mass}
The temperature dependence of the pseudoscalar mass has been parameterized with a quartic polynomial in $\hat{g}^2$ of the form
\begin{align}
\label{eq:quartic}
    \frac{m_P}{2\pi T} = p_0 + p_2 \hat{g}^2 + p_3 \hat{g}^3 + p_4 \hat{g}^4 \, .
\end{align}
The leading and the quadratic coefficients have been found to be compatible with the free theory value and the next-to-leading order correction, in eq. \eqref{eq:meson_pt}, respectively. Once $p_0$ and $p_2$ have been fixed to their corresponding perturbative values, we obtain for the cubic and the quartic fit parameters $p_3=0.0038(22)$ and $p_4=-0.0161(17)$ with ${\rm cov}(p_3,p_4)/[\sigma(p_3)\sigma(p_4)]=-1.0$ with an excellent $\chi^2/{\rm dof} =0.75$. Such a polynomial is displayed, as a red band, together with the non-perturbative data in the left panel of figure \ref{fig:massCL}. It is clear that the quartic term is necessery to explain the behaviour of the non-perturbative data in the entire range of temperature. In particular, at the highest temperatures it contributes for about $50\%$ of the total contribution induced by interactions, while at low temperature it competes with the quadratic coefficient to bend down the pseudoscalar mass.
\subsubsection{Vector mass}
The mass difference between the vector and the pseudoscalar mass is due to spin-dependent contributions, which, as we have anticipated, are expected to be $O({g}^4)$ in the effective field theory. By plotting our results as a function of $\hat{g}^4$, see right panel of figure \ref{fig:massCL}, these turn out to lie on a straight line with vanishing intercept in the entire range of temperature. We then parameterized the temperature dependence with
\begin{align}
    \label{eq:spindep}
    \frac{(m_{V} - m_{P})}{2\pi T} = s_4\, \hat g^4\;  
\end{align}
and we obtain $s_4=0.00704(14)$ with $\chi^2/{\rm dof}=0.79$. It is remarkable that, even at the highest temperatures which was simulated, the mass difference is clearly different from zero within the statistical error, a fact which is not expected by the next-to-leading order estimate in eq. \eqref{eq:meson_pt}, obtained in the effective field theory. Then the best parameterization for the vector screening mass is given by
\begin{align}
    \label{eq:mvfit}
    \frac{m_{V}}{2\pi T} = p_0 + p_2\, \hat g^2 + p_3\, \hat g^3 + (p_4 + s_4) \, \hat g^4\; ,
\end{align}
with covariances ${\rm cov}(p_3,s_4)/[\sigma(p_3)\sigma(p_4)]=0.08$ and ${\rm cov}(p_4,s_4)/[\sigma(p_4)\sigma(p_4)]=-0.07$. In the vector channel the quartic contribution appearing in eq. \eqref{eq:mvfit} is responsible for about $15\%$ of the total contribution due to interaction at the electro-weak scale. Moreover, the quartic coefficient for the vector screening mass is about $50\%$ smaller than the corresponding coefficient for the pseudoscalar channel. As a consequence, its contribution is not large enough to compete with the quadratic coefficient and to bend down the value of the vector mass at low temperature. For a more detailed analysis of the results we refer to section 7 of Ref. \cite{DallaBrida:2021ddx}.
\subsection{Baryonic screening masses}
\begin{wrapfigure}{r}{0.5\textwidth}
\centering
    \includegraphics[width=\linewidth]{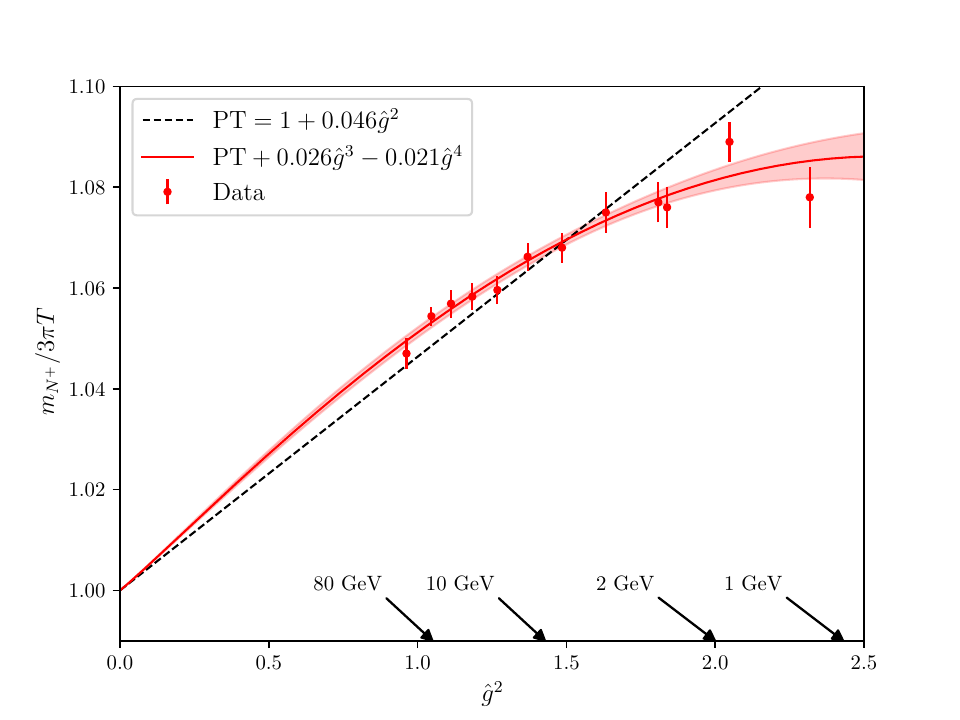}
    \caption{Nucleon screening mass versus $\hat g^2$. The band      represent the best fit to eq.~(\ref{eq:quartic-baryon}), while the dashed line is the analytically known contribution in eq.~\eqref{eq:baryon_pt}.\label{fig:mass-baryon-CL}}
\end{wrapfigure}

In contrast with the mesonic case, there are very few studies on the baryonic sector both on the lattice \cite{Datta:2012fz,Rohrhofer:2019yko,Aoki:2020noz} and in the three dimensional effective theory \cite{Hansson:1994nb} and for what concerns lattice calculations, no continuum limit extrapolation has ever been performed.
In Ref. \cite{Giusti:2024ohu} we computed the baryonic screening masses for the first time with continuum limit extrapolations and with a final accuracy of a few permille from $1$ GeV up to the electro-weak scale. As expected in a chirally symmetric regime, the positive and the negative parity screening masses are found to be degenerate in the entire range of temperatures. For this reason, in the following we only focus on the positive parity mass $m_{N^+}$. The final results are shown in figure \ref{fig:mass-baryon-CL} as a function of $\hat{g}^2(T)$, see eq. \eqref{eq:gmu}. 

As it is clear from the plot, the bulk of the baryonic screening mass is given by the free field theory $3\pi T$ plus a $4-8\%$ positive contribution due to interaction. It is rather clear that
from $T\sim160$ GeV down to $T\sim5$ GeV the perturbative expression is within half a percent with respect to
the non-perturbative data. The full set of data, however, shows a distinct negative curvature which requires higher orders in the coupling constant to be parameterized. Similarly to the case of the mesonic screening masses, the temperature dependence of the baryonic screening mass has been parameterized with the ansatz
\begin{align}
    \label{eq:quartic-baryon}
\frac{m_{N^+}}{3\pi T} = b_0 + b_2\, \hat g^2 + b_3\, \hat g^3 + b_4\, \hat g^4\; .  
\end{align}
$b_0$ and $b_2$ turn out to be compatible with the free-theory and the next-to-leading values in eq. \eqref{eq:baryon_pt} respectively. Then, by enforcing those values and fitting again, we obtain $b_3=0.026(4)$, $b_4=-0.021(3)$ and ${\rm cov}(b_3,b_4)/[\sigma(b_3) \sigma(b_4)]=-0.99$ with $\chi^2/{\rm dof}= 0.64$, which is the best parameterization of our results over the entire range of temperatures explored. Notice that, in general, other parameterizations of the lattice data are possible as well. These, however, result in the disagreement between the fit parameter $b_2$ and the 1-loop perturbative correction in eq. \eqref{eq:baryon_pt}. For a more detailed discussion on such parameterizations we refer to section 5 of Ref. \cite{Giusti:2024ohu}.
\label{sec:baryon}

%% file: conclusion.tex
In these proceedings we outlined a recently proposed strategy to simulate QCD at very high temperatures on the lattice with a moderate computational effort \cite{DallaBrida:2021ddx}. This strategy combines the use of a non-perturbative definition of the running coupling in a finite volume to renormalize the theory and the use of shifted boundary conditions, as well as the steady, theoretical and algorithmic progress in simulation of gauge theory which makes it possible to simulate large lattices with a moderate computational effort.

As a first application of this strategy, we computed both the mesonic \cite{DallaBrida:2021ddx} and the baryonic \cite{Giusti:2024ohu} screening masses from $1$ GeV up to the electro-weak scale. In both cases, we obtained results with a few permille accuracy in the continuum limit. By scrutinizing in detail the temperature dependence of these masses, we found that the next-to-leading perturbative calculation, both in the mesonic and in the baryonic sector, is not sufficient to explain the behaviour of our data, since higher order contributions in the running coupling remain relevant in the entire range of temperatures. The results presented at this conference pave the way for a fully non-perturbative treatment of thermal QCD up to the electro-weak scale and shed new light on the poor convergence of the perturbative approach at very high temperatures.

%% file: ack.tex
\\[2mm]\noindent\textbf{\large{Acknowledgements }} We acknowledge PRACE for awarding us access to the HPC system MareNostrum4 at the Barcelona Supercomputing Center (Proposals n. 2018194651 and 2021240051) and EuroHPC for the access to the HPC system Vega (Proposal n. EHPC-REG-2022R02-233) where most of the numerical results presented in this paper have been obtained. We also thank CINECA for providing us with computer time on Marconi (CINECA-INFN, CINECA-Bicocca agreements). The R\&D has been carried out on the PC clusters
Wilson and Knuth at Milano-Bicocca. We thank all these institutions for the technical support. This work is (partially) supported by ICSC – Centro Nazionale di Ricerca in High Performance Computing, Big Data and Quantum Computing, funded by European Union – NextGenerationEU.

%% file: appA.tex
\label{app:A}
In this appendix we sketch the perturbative calculation for the baryonic screening mass which leads to the result in eq. \eqref{eq:baryon_pt}. Such a calculation is performed in the effective field theory described in section \ref{sec:eft} in which quarks are treated as heavy, static fields interacting with soft and ultrasoft gauge modes only. For a more detailed discussion see Ref. \cite{Giusti:2024mwq}.
\subsection{Equations of motion and quark propagators at next-to-leading order}
Let us introduce the quark propagators for the $\chi$ and $\phi$ fields defined in eq. \eqref{eq:chiphi} as
\begin{align}
    S_\chi(x) \equiv \expval{\chi(x) \bar\chi(0)}_f \, , \qquad S_\phi(x) \equiv \expval{\phi(x) \bar\phi(0)}_f \,,
\end{align}
where $\expval{\cdot}_f$ refers to the fact that the expectation value is taken by integrating over the fermionic variables in the path integral. From the effective action in eq. \eqref{eq:NRQCD} it is straightforward to see that such propagators satisfy the equations of motion
\begin{align}
\label{eq:S_chi}
 & \expval{\left[ M + \partial_3^{ }   - \frac{\nabla_\perp^2}{2\pi T}\right]
 S_{\chi}^{ } (x)}= 
 g_\rmii{E}^{ } \expval{\Big[i A_3^{ }(x) + A_0^{ }(x)\Big]
 S_{\chi}^{ } (x)} - i \id \delta^{(3)}_{ }(x)\, ,\\[0.125cm]
 & \expval{\left[ M + \partial_3^{ }
  - \frac{\nabla_\perp^2}{2\pi T}\right] S_\phi^{ } (x)}=
 g_\rmii{E}^{ } \expval{\Big[i A_3^{ }(x) - A_0^{ }(x)\Big] S_\phi^{ } (x)}
 + i \id \delta^{(3)}_{ }(x) \, ,
\label{eq:S_phi}
\end{align}
where $\id$
stands for the identity in spinor and colour indices.  
Since the fermions have been integrated out, 
the expectation values in eqs.~(\ref{eq:S_chi}) and (\ref{eq:S_phi})
indicate the path integral over the gauge fields. 
Note that these equations are valid
also without integrating over the gauge fields,
i.e.\  for a fixed gauge field background, and that
at this order the fermion
propagators are diagonal in flavour and spin.
The equations of motion above can be solved perturbatively, at next-to-leading order, by writing
\begin{align}
 S_{\chi}^{ } (\mathbf{r}, x_3^{ }) =
  S_{\chi}^{(0)} (\mathbf{r}, x_3^{ })
 + g_\rmii{E}^{ }\, S_{\chi}^{(1)} (\mathbf{r}, x_3^{ }) 
 + O(g_\rmii{E}^2 )\; , 
\end{align}
and analogously for $S_{\phi}^{ } (\mathbf{r}, x_3^{ })$, where, by inserting these expressions in the equations of motion, at leading order we obtain
\begin{align}
    \label{eq:free_prop1}
    S_{\chi}^{(0)} (\mathbf{r}, x_3^{ }) 
  = -i  \theta(x_3^{ }) \id\!\! \int_{\textbf{p}}
  e^{i \textbf{p} \cdot \textbf{r}  }\, 
   e^{-x_3 \bigl( M +\frac{\textbf{p}^2}{2 \pi T} \bigr)} \, ,\qquad
 S^{(0)}_\phi (\mathbf{r}, x_3^{ }) 
 = - \, S^{(0)}_\chi (\mathbf{r}, x_3^{ })\, ,     
\end{align}
where $\int_{\mathbf{p}}\equiv \int\! {\rm d }^2_{ } \mathbf{p}/(2\pi)^2_{ }$. While at next-to-leading, the contributions to the quark propagators can be written in terms of the leading contributions as
\begin{align}
\label{eq:ge_prop1}
 S_{\chi}^{(1)} (\mathbf{r}, x_3^{ })
 & \,\simeq\, \int_0^{x_3}\! {\rm d}z^{ }_3\, \big[i A_3^{ } + A_0^{ }\big]
 \Big(\frac{z_3}{x_3}\textbf{r},z_3^{ }\Big)
 \, S_{\chi}^{(0)} (\mathbf{r}, x_3^{ })
 \,, \\
 S_{\phi}^{(1)} (\mathbf{r}, x_3^{ })
 & \,\simeq\, \int_0^{x_3}\! {\rm d}z^{ }_3\, \big[i A_3^{ } - A_0^{ } \big]
\Big(\frac{z_3}{x_3}\textbf{r},z_3^{ }\Big)\,
 S_{\phi}^{(0)} (\mathbf{r}, x_3^{ })
 \,,
\label{eq:ge_prop2}
\end{align} 
where we assumed heavy quarks in approximating the motion in the transverse directions.
\subsection{Baryonic correlators in the effective theory}
In the effective field thoery, the nucleon interpolating operator in eq. \eqref{eq:nucleon_op} is readily obtained by using the definitions in appendix A.2 of Ref. \cite{Giusti:2024mwq}. By assuming the baryon to propagate in the positive $x_3$-direction and by displacing each quark field in the transverse ($x_1,x_2$)-direction, the operator can be written in the three-dimensional non-relativistic effective theory as
\begin{align}
N (\mathbf{r}_1^{ },\mathbf{r}_2^{ },\mathbf{r}_3^{ };x_3^{ })
 & \to
 \epsilon^{abc}_{ }\, 
 \big[\, 
 \chi^{aT}_{u} (\mathbf{r}_1^{ },x_3^{ })
 \,\sigma_2^{ }\, \phi_{d}^b(\mathbf{r}_2^{ },x_3^{ })
 + \phi^{aT}_{u}(\mathbf{r}_1^{ },x_3^{ })
 \,\sigma_2^{ }\, \chi_{d}^b(\mathbf{r}_2^{ },x_3^{ })
 \,\big]
 \, \chi^{c}_{d,\alpha}(\mathbf{r}_3^{ },x_3^{ })
 \,, \nonumber \\[0.25cm]
 \overline{\!N} (0)
 & \to  
 \epsilon^{feg}_{ }\,
 \big[\, 
 \bar\phi^{f}_{d}({0}^{ })
 \,\sigma_2^{ }\,
 \bar\chi_{u}^{gT}({0}^{ })
  + 
 \bar\chi^{f}_{d}({0}^{ })
 \,\sigma_2^{ }\, 
 \bar\phi^{g T}_{u}({0}^{ })
 \,\big]
 \, 
 \bar\chi^{e}_{d,\alpha}({0}^{ }) 
  \,,
 \label{eq:N_FT}
\end{align}
where $\alpha$ is a two-component spinor index. 
The nucleon two-point correlators (see eq.~\eqref{eq:2pt-baryon} with $\xi=0$) are defined
in the effective theory as
\begin{align}
  \label{eq:Ncorr_eff}
   {\cal C}_{\pm}^{ }
   (\textbf{r}_1^{ },\textbf{r}_2^{ },\textbf{r}_3^{ };x_3^{ })
    &\equiv
   \frac{1}{T} \, 
   \Tr \expval{N(\textbf{r}_1^{ },\textbf{r}_2^{ },\textbf{r}_3^{ };x_3^{ })
    \,\overline{\!N}(0) P_{\pm}^{ }} \,\\
    &=\mp \, T^2_{ } \,\Big\langle\,
   2 W(\textbf{r}_1^{ },\textbf{r}_2^{ },\textbf{r}_3^{ };x_3^{ })
 + 3 W(\textbf{r}_2^{ },\textbf{r}_1^{ },\textbf{r}_3^{ };x_3^{ }) 
        \,\Big\rangle \,,
\label{eq:Ncorr_eff_wick}
\end{align}
where
$
 P_{\pm}^{ } 
 =
 (\pm i/2) \id^{ }_{ }
 \leftrightarrow
 [\gamma^{ }_0 
 (\id^{ }_{ } \pm \gamma^{ }_3)/2]^{ }_{11}
$ and in the second line we exploited the antisymmetry of the Levi-Civita symbol and we performed the integration over the fermionic fields.
The Wick contraction above is defined in terms of quark propagators as
\begin{align}
 \label{eq:Wick}
 W(\textbf{r}_1^{ },\textbf{r}_2^{ },\textbf{r}_3^{ };x_3^{ })
  \; \equiv \; 
 -i  \, \epsilon^{abc}_{ } \epsilon^{gfe}_{ }\, 
 S^{ag}_{\chi} (\mathbf{r}_1^{ }, x_3^{ })\,
 S^{bf}_{\phi} (\mathbf{r}_2^{ }, x_3^{ }) 
 S^{ce}_{\chi} (\mathbf{r}_3^{ }, x_3^{ })\; .
\end{align}
Given that the two Wick contractions 
in \eqref{eq:Ncorr_eff_wick} differ just by a permutation
of coordinates, and that in the end all coordinates are set equal
(cf.\ eq.~\eqref{eq:Ncorr_eff} and \eqref{eq:2pt-baryon}), these 
yield the same baryonic screening mass. 
\subsection{The Schr\"odinger equation}
Once the expression of the Wick contraction in eq. \eqref{eq:Wick} is known, its equation of motion at order $O(g^2)$ in the strong coupling constant is readily worked out from the equations of motion in eq. \eqref{eq:S_chi} and \eqref{eq:S_phi}. By performing the gluon contractions and by taking the large separation limit in the $x_3$-direction, since we are interested in extracting the screening mass, the equation of motion for a generic two-point correlation function associated with a baryonic interpolating operator reads
\begin{align}
\label{eq:schrodinger}
    \left[ \partial_3 -\sum_{i=1}^3 \frac{\nabla_{\mathbf{r}_i}^2}{2\pi T} + V(\mathbf{r}_1, \mathbf{r}_2, \mathbf{r}_3) \right] \expval{W(\mathbf{r}_1,\mathbf{r}_2,\mathbf{r}_3;x_3)} = 0 + O(g^3) \,,
\end{align}
with
\begin{align}
    V(\mathbf{r}_1, \mathbf{r}_2, \mathbf{r}_3) \equiv 3M + \frac{1}{2}\left[ V^-(r_{12}) + V^+(r_{13}) + V^- (r_{23}) \right] \,,
\end{align}
where $\mathbf{r}_{ij}=|r_i-r_j|$ and we introduced the static potentials $V^\pm(r)$, defined in Ref. \cite{Brandt:2014uda},
see Ref. \cite{Giusti:2024mwq} for the details.
Notice that eq. \eqref{eq:schrodinger} is simply a (2+1)-dimensional Schr\"odinger equation and the screening mass associated to the two-point correlation function in eq. \eqref{eq:Ncorr_eff_wick} is obtained by extracting the lowest energy eigenvalue of such a system.

The numerical solution of the Schr\"odinger equation has been carried out in two different and independent ways, involving a two dimensional generalization of the hyperspherical harmonics method, see appendix D of Ref. \cite{Giusti:2024mwq}, and by discretizing the Hamiltonian on a mesh grid. Both ways return the value, up to all digits shown, which is reported in eq. \eqref{eq:baryon_pt}.